\documentclass{mem}
\usepackage{natbib}\usepackage{txfonts}\usepackage{balance}
\usepackage{graphicx}
\usepackage[a4paper]{hyperref}
\idline{75}{1}

\begin{document}
\def\teff{$T\rm_{eff }$}
\def\kms{$\mathrm {km s}^{-1}$}

\title{
The investigation of particle acceleration in colliding-wind massive binaries with SIMBOL-X.}
   \subtitle{}

\author{
M. \,De Becker\inst{1} 
\and 
G. \,Rauw\inst{1}
\and 
J.M. \,Pittard\inst{2}
\and
R. \,Blomme\inst{3}
\and
G.E. \,Romero\inst{4}
\and
H. \,Sana\inst{5}
\and
I.R. \,Stevens\inst{6}
          }

  \offprints{M. De Becker}

\institute{Institut d'Astrophysique et de G\'eophysique, Universit\'e de Li\`ege, FNRS, Belgium\\
	\email{debecker@astro.ulg.ac.be}
	\and
	School of Physics and Astronomy, The University of Leeds, UK
	\and
	Royal Observatory of Belgium, Brussels, Belgium
	\and
	Facultad de Ciencias Astronom\'{\i}cas y Geof\'{\i}sicas, Universidad Nacional de La Plata, Argentina
	\and
	European Southern Observatory, Chile
	\and
	School of Physics and Astronomy, University of Birmingham, UK
}

\authorrunning{De Becker et al.}

\titlerunning{Particle acceleration in colliding-wind binaries}

\abstract{An increasing number of early-type (O and Wolf-Rayet) colliding wind binaries (CWBs) is known to accelerate particles up to relativistic energies. In this context, non-thermal emission processes such as inverse Compton (IC) scattering are expected to produce a high energy spectrum, in addition to the strong thermal emission from the shock-heated plasma. SIMBOL-X will be the ideal observatory to investigate the hard X-ray spectrum (above 10 keV) of these systems, i.e. where it is no longer dominated by the thermal emission. Such observations are strongly needed to constrain the models aimed at understanding the physics of particle acceleration in CWB. Such systems are important laboratories for investigating the underlying physics of particle acceleration at high Mach number shocks, and probe a different region of parameter space than studies of supernova remnants.
}
\maketitle{}

\section{Colliding-wind binaries as particle accelerators}
\begin{figure}[t!]
\resizebox{\hsize}{!}{\includegraphics[clip=true]{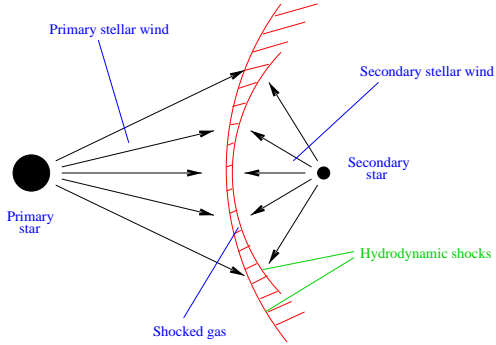}}
\caption{\footnotesize
Schematic view of a CWB system.
}
\label{cwb}
\end{figure}

More than 30 early-type stars (most of them confirmed or suspected binaries) exhibit synchrotron radiation in the radio domain \citep[see][]{thesis}. This indicates that relativistic electrons are present, and therefore that a particle acceleration process is at work. The acceleration mechanism is most probably Diffusive Shock Acceleration (DSA) in the presence of strong hydrodynamic shocks \citep[e.g.][]{PD140art}. In massive CWBs, particle acceleration is expected to occur at the global shocks bounding the wind-wind collision region, and perhaps also within this volume (see Fig.\,\ref{cwb}). Such systems are further considered to be candidates for the production of cosmic-rays. In the context of CWBs, non-thermal X-rays are also expected to be produced. Previous observations with XMM suggested hints for IC emission \citep{thesis}, although INTEGRAL data provided only upper limits on the hard X-ray flux of CWBs \citep{cygint}.

\section{Cyg\,OB2\,\#8A as a test case}
XMM-Newton observations of this O6If + O5.5III(f) binary (P\,$\sim$\,22\,d) revealed X-ray emission ($<$\,10\,keV) dominated by the plasma heated ($\sim$\,20\,MK) by the colliding winds \citep{DeBcyg8a}. Any putative non-thermal emission component due to IC scattering is most probably overwhelmed by the thermal emission. The hard X-ray domain ($>$\,10\,keV) needs to be investigated in order to study the non-thermal emission from CWB.

\begin{figure}[t!]
\resizebox{\hsize}{!}{\includegraphics[clip=true]{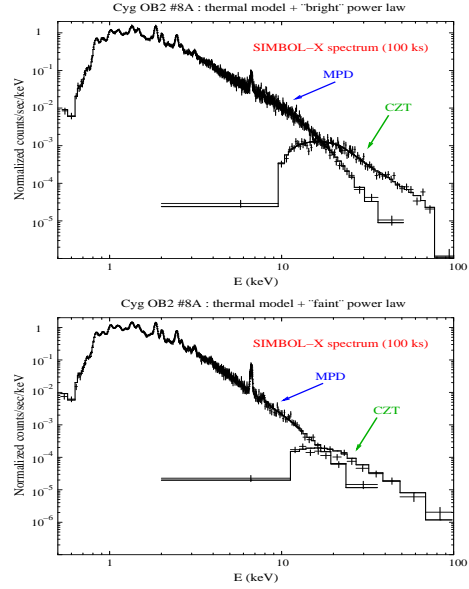}}
\caption{\footnotesize
Synthetic SIMBOL-X MPD and CZT spectra of Cyg\,OB2\,\#8A for two different assumptions on the flux of the power law.
}
\label{simxspec}
\end{figure}

We simulated SIMBOL-X spectra on the basis of a model combining the thermal emission (see XMM results) and a power law (IC scattering) with a photon index equal to 2. This latter value is typical of relativistic electrons strongly affected by IC scattering that steepens the power law. We considered a "bright" case with a flux a factor of 10 lower that the upper limits derived by INTEGRAL-ISGRI observations \citep{cygint}, and a "faint" case about a factor 10 lower again. The SIMBOL-X synthetic spectra obtained in both cases with an exposure time of 100\,ks are shown in Fig.\,\ref{simxspec}. The quality of these spectra should allow us to derive the flux and the photon index of the power law. Ideally, observations at different orbtial phases would be taken in order to study the variation of the non-thermal emission along the eccentric orbit.

\section{Conclusions}
Our objective is to understand and quantify the particle acceleration process in massive CWB and its potential relation with the production of cosmic-rays, by putting constraints upon the high-energy emission (photon index, flux, phase-locked variability) from such systems.

\begin{acknowledgements}
MD is grateful to Giusi Micela for her kind invitation to participate in this fruitful workshop.
\end{acknowledgements}

\bibliographystyle{aa}

\end{document}